\documentclass[11pt, a4paper]{article}
\usepackage{jheppub}
\usepackage{graphicx}
\usepackage{epsfig,amsmath,amsthm}

\newcommand{\be}{\begin{equation}}
\newcommand{\ee}{\end{equation}}
\newcommand{\bea}{\begin{eqnarray}}
\newcommand{\eea}{\end{eqnarray}}
\def\bse{\begin{subequations}}
\def\ese{\end{subequations}}
\newcommand{\IR}{\mathbb{R}}

\def\IZ{\relax\ifmmode\hbox{Z\kern-.4em Z}\else{Z\kern-.4em Z}\fi}

\newcommand{\non}{\nonumber \\}

\def\half{\frac{1}{2}} 

\def\del{{\partial}}

\def\presub{\vspace{.5cm} \noindent}

\def\bi{\begin{itemize}} \def\ei{\end{itemize}}

\def\({\left(} \def\){\right)}
\def\[{\left[} \def\]{\right]}
\title{\center{The Fields of Ultra-Relativistic Gravitation}}

\author{Barak Kol\\
 {\it Racah Institute of Physics, Hebrew University\\
 Jerusalem 91904, Israel}\\
 {\tt\href{mailto:barak_kol@phys.huji.ac.il}{barak\_kol@phys.huji.ac.il}}
 }

\abstract{In order to facilitate the study of weak
ultra-relativistic (and Planckian) scattering we present an
appropriate decomposition of the gravitational field together with
its whole non-linear action. More generally the results apply to
any reduction over a non-degenerate spacetime fiber.}
\begin{document}

\maketitle

\section{Introduction}

In a companion paper we discuss an effective field theory for weak
ultra-relativistic scattering. The problem is to analyze the
scattering of two light-like (or at least ultra-relativistic)
point-like particles at a large impact parameter (see
\cite{planck} and references therein). One finds that the physics
localizes to the moment of passing in the longitudinal direction
and interactions take place only in the transverse space.

When the two particles interact through the gravitational
interaction, the components of the gravitational field split into
several fields from the transverse perspective. In the literature
(see for example \cite{ACV93}) the gravitational Einstein-Hilbert
action is expanded perturbatively for weak gravitational fields.
In this paper we shall define the field decomposition
(\ref{decomposition}) and compute its whole non-linear action
(\ref{URG-action}). Expanding it allows to readily read off all
the gravitational bulk propagators and vertices for the
perturbation theory. For example we reproduce a leading propagator
term in (\ref{Hprop}).

In the related case of the post-Newtonian approximation (see
\cite{BlanchetRev,Schaefer-account} for reviews, and
\cite{GoldbergerRothstein1} for an effective field theory
approach) it is natural and useful to decompose the gravitational
field through a temporal Kaluza-Klein reduction into
Non-Relativistic Gravitational (NRG) fields
\cite{CLEFT-caged,NRG}. These consist of the Newtonian potential,
the gravito-magnetic 3-vector and a spatial metric. The full,
non-linear gravitational action for these fields was determined in
\cite{NRGaction}.

Here we go further by allowing dimensions larger than one (and
arbitrary in principle) for both (the longitudinal) fiber and (the
transverse) base. Indeed, our main expression (\ref{URG-action})
includes novel terms which were not all present in the earlier
cases. \footnote{I was notified that an action whose mathematical form is essentially the same was given by Yoon \cite{Yoon}.
 While the mathematical context and tools there are essentially the same as here, namely a Kaluza-Klein reduction, the physical context and application is completely different and has no relation to ultra-relativistic gravitation. Instead \cite{Yoon} interprets 4d General Relativity as a 1+1 gauge theory. Technically, there the $1+1$ space is the base while here it is a fiber, and here spacetime dimension is arbitrary.
 I thank S. Carlip for alerting me to \cite{Yoon}.}

Computing the action by the standard method metric $\to$
Christoffel symbols $\to$ curvature tensor $\to$ action would be a
very complicated analytical task, perhaps hopelessly so. Here we
simplify the computation to manageable form with no computerized
computation by using a non-orthonormal frame within Cartan's
method, namely a hybrid method which incorporates both a
non-trivial frame and a non-trivial metric as in \cite{NRGaction}.

\section{Field decomposition and action}

We work in the center of mass frame of a $d$ dimensional spacetime
and we denote the longitudinal direction by $z$, and the
transverse directions by $x^i$.

In the leading ultra-relativistic limit transverse gradients
dominate over longitudinal ones \be
 \del_a \equiv \del_z \ll \del_i \equiv \del_x \ee
%the sources and accordingly the propagator of the gravitational field is independent of the longitudinal coordinates $z,t$ (or equivalently $z^\pm$).
Therefore it is natural to perform a dimensional reduction \`{a}
la Kaluza-Klein (KK) \cite{Kaluza-Klein} of the metric over the
light-cone coordinates $z^\pm$, a reduction which highlights the
transformation properties (or tensor nature) with respect to gauge
transformations which depend only on the transverse directions $x$
\bea
 g_{\mu\nu} &\to& \(G_{ab},\, A^a_i,\, g_{ij}\) \non
 ds^2 &=& G_{ab} ( dz^a-A^a_i dx^i ) ( dz^b-A^b_j dx^j ) - g_{ij} dx^i dx^j  \label{decomposition}
 \eea
 In this expression $a,b=+,-$ and $i,j=1,2,\dots,d-2$. Note that
 this dimensional reduction is more general and it applies to a
reduction to a any base manifold $X$ parameterized by $x^i$ over a
any fiber $Z$ (not necessarily 2d) parameterized by the
coordinates $z^a$. We observe that from the transverse perspective
the fields are $G_{ab}$ a symmetric matrix of 3 scalars, $A^a_i$ a
pair of transverse vectors, and $g_{ij}$ the transverse metric. To
put this dimensional reduction in the context of the literature we
recall that the original and standard KK reduction
\cite{Kaluza-Klein} is over spatial directions, the
Non-Relativistic Gravitational reduction (NRG)
\cite{CLEFT-caged,NRG} is over the time direction, while here the
reduction is over a Lorentzian $1+1$ fiber.

The action is simply \bea
 S &=& \frac{1}{16 \pi G} \int \sqrt{-G} \sqrt{g}
d^{d-2}x d^2z  \cdot \non & &
 \left\{- \left<
K_{abi}[G]\right>_{dW}^2  + \frac{1}{4} \bar{F}^2 + \bar{R}[g]
\right. %\non & &
 \left. +\left< \half \del_a g_{ij} \right>_{dW}^2 - R[G] \right\}
 \label{URG-action}
\eea

We proceed to define all the symbols and conventions. First \bea
 G &:=& \det G_{ab} \non
 g &:=& \det g_{ij} \\
 D_i &=& \del_i  + A^a_i \del_a
 \eea

The extrinsic curvature of the $1+1$ fiber is
 \footnote{In general the extrinsic curvature $K$ evaluated on two vector fields $X,Y$ which lie in a sub-manifold
 is defined by $K(X,Y)=\(D_X Y \)^\perp$. This defines a symmetric tensor.}
% It is symmetric for torsion free connections. $\[X,Y\]^\perp=0$.
% K_{ab}^i is essentially \Gamma_{ab}^i, which is given by omega^i_{ab} in a non-coordinate basis
% recall the definition of the connection omega: d e_A = e_B \omega^A_B (MTW 14.14 p. 350)
\be
 K_{abi}[G] := - \half \( D_i G_{ab}+ G_{c(a} \del_{b)} A^c_i \) = - \half \( \del_i G_{ab}+ {\cal L}_{A_i}
 G_{ab} \)
 \ee
 where in the last expression ${\cal L}_{A_i}$ denotes the Lie
derivative with respect to the longitudinal vector $A_i \equiv
A_i^a$.
 \footnote{Recall that the Lie derivative by a vector field
$V$ of a vector field $W$ is defined as ${\cal L}_V W^\mu := \[
V,W \] := V^\nu \del_\nu W^\mu - W^\nu \del_\nu V^\mu$ where $\[
V,W \]$ denotes the commutator of the two vector fields. When one
extends this derivation to all tensors one finds that the Lie
derivative of a co-vector $\omega$ is given by ${\cal L}_V
\omega_\mu = V^\nu \del_\nu \omega_\mu + \omega^\nu \del_\mu
V^\nu$. Similarly ${\cal L}_V G_{ab} = V^c \del_c G_{ab} + G_{ac}
\del_b V^c + G_{bc} \del_a V^c$.}
 The $(-1/2)$ prefactor was inserted to conform with the
standard definition of the extrinsic curvature.
% K_{abi}=\Gamma_{iab} - christoffel symbol

For any symmetric tensor field we define a ``deWitt'' quadratic
form (actually once applied to a differential of a metric and
integrated over the manifold it becomes a metric on the space of
metrics \cite{deWitt-metric}, see \cite{NRGaction} for its
appearance in the NRG action) \be
 \left< h_{IJ}\right>_{dW}^2 := \left| h_{IJ} \right|^2 - h^2
 \ee
 In particular \bea
 \left< K_{abi}[G]\right>_{dW}^2 &=&   g^{ij} G^{ac} G^{bd}\, \( K_{abi} K_{cdj} - K_{aci} K_{bdj} \) \non
 \left< \half \del_a g_{ij} \right>_{dW}^2 &=& \frac{1}{4} G^{ab}\, g^{ik} g^{jl} \[  \del_a g_{ij}\, \del_b g_{kl} - \del_a g_{ik}\, \del_b g_{jl} \] \eea

The generalized (magnetic) field strength\footnote{Note that as
usual the $1/4$ prefactor could have been avoided had we
accompanied the anti-symmetrization in the definition of $\bar{F}$
with a division by $2$.}
 is defined by
\be
  \bar{F}_{ij}^a := D_i A_j^a - D_j A_i^a = F_{ij}^a + A_i^b \del_b A_j^a -
  A_j^b \del_b A_i^a \label{def-barF}
\ee
 and its square is given by \be
\bar{F}^2 = G_{ab} g^{ik} g^{jl}\, F_{ij}^a F_{kl}^b ~.
 \ee

Finally $\bar{R}[g]$ denotes the Ricci scalar of the transverse
metric $g$ where the derivatives in its expression are replaced
everywhere as follows $\del_i \to D_i$. Borrowing notation from
the {\it Mathematica} software this definition can be stated by
\be
 \bar{R}[g] := R[g] ~ /. ~ \del_i \to D_i ~.\ee

\presub {\bf Derivation}. In order to compute the action we used a
non-orthonormal frame within Cartan's method, namely a hybrid
method which incorporates both a non-trivial frame and a
non-trivial metric as in \cite{NRGaction}. This action generalizes
an analogous result from the KK literature found by Aulakh and
Sahdev \cite{AulakhSahdev}, and the NRG (Non-Relativistic
Gravitational) action \cite{NRGaction}.

\presub {\bf Tests}. We tested the Ultra-Relativistic
Gravitational (URG) action (\ref{URG-action}) in several limits.
For $d_Z=1$ we reproduce the KK \cite{AulakhSahdev} and NRG
actions.
% S = \frac{1}{16 \pi G} \int \frac{1}{4} F^2 + R[g] + \< \half \dot{g} \>^2_{dW}
For $d_X=1$ we reproduce the ADM \cite{ADM} action (see for
example \cite{NRG-ADM}).
% With A \to -A
% S = \frac{1}{16 \pi G} \int R[g] + \< K[g] \>^2_{dW}
As a Final test for $A^a_i=0$ the action is symmetric with respect
to the exchange $X \leftrightarrow Z$.
% Actually once we replace g \to -g the action becomes symmetric:
% S = \frac{1}{16 \pi G} \int -R[g] -R[G] + \< \half \del_a g \>^2_{dW} + \< \half \del_i G] \>^2_{dW}
% These properties might be enough to fix the action
The ``stationary limit'', namely no $z$ dependence, is another
interesting limit. In this limit the last two terms in the action
(\ref{URG-action}) vanish,\footnote{If a curved $Z$ fiber is
allowed then some $R[G]$ would remain.} while the other terms
simplify: $D_i \to \del_i, ~ \bar{R} \to R$, $\bar{F} \to F$ and
$K_{abi} \to -(1/2) \del_i G_{ab}$.

\section{Perturbing around flat longitudinal spacetime}

The dimensionally reduced action can be used to linearize the
action around any prescribed product space-time $X \times Z$. In
our case the unperturbed space-time is flat and accordingly we may
use \bea
 G_{ab} &=& \eta_{ab} + H_{ab} \non
 g_{ij} &=& \delta_{ij} + h_{ij} ~. \eea

At leading ultra-relativistic order one source couple dominantly
to $H^{++}$ while the others couples to $H^{--}$. For 2d metrics
the deWitt metric simplifies \be
 \left< G_{ab} \right>^2_{dW} = \frac{2}{-G}  \( dH_{++} dH_{--} - dH_0^2 \)\non
\ee
 where \bea
 H_0 &:=& H_{+-} \non
 -G &=& (1+H_0)^2-H_{++} H_{--} ~.
 \eea
% direct computation on Mathematica - there could be a way to show it from symmetry
Substituting into (\ref{URG-action}) we find that the propagator
for $H_{++}, \, H_{--}$ is  \be
 S \supset -\frac{1}{32 \pi G} \int d^{d-2}x\, d^2z\, \vec{\nabla}_\perp
 H_{++}\, \vec{\nabla}_\perp H_{--} ~. \label{Hprop} \ee

Longitudinal boosts are a global $SO(1,1) \simeq \IR$
 symmetry of the action.\footnote{From the
transverse point of view $G_{ab}$ are three scalars and being a
quadratic form the action is invariant under a similarity
transformation $G \to R^T G R$ for any $R \in GL(2,\IR)$. Moreover
the vacuum (unperturbed solution) $G_{ab}=\eta_{ab}$ is invariant
under the $SO(1,1) \simeq \IR$ subgroup of 2d Lorentz
transformations, which accordingly is a symmetry of the linearized
action.} Under this symmetry $H_{++}$ has charge $+2$ while
$H_{--}$ has charge $(-2)$. This symmetry can be represented in
the Feynman rules by representing the fields $H_{++},\, H_{--}$ by
an oriented line and distinguishing them by its orientation, which
represents the flow of charge.

\presub {\bf Some open questions}. \bi
 \item In 4d spacetime the transverse metric
is 2d which may bring about additional simplifications.
% complex fields?? In ACV93 they replace the 2d transverse metric g_{ij} by a complex non-local field phi
% according to \del_*^2 \phi = h_{\mu\nu} \eta^\mu \eta^\nu
% Our notation is opposite to theirs: they denote by z the transverse direction and with x the longitudinal ones
 \item In the post-Newtonian case Weyl rescaling of the metric was employed, and actually was necessary even to reproduce the Newtonian potential.
In 4d space the transverse space is 2d and Weyl rescaling is less
effective, but it could possibly be of use at least in higher
dimensions.
 \ei

\end{document}